\documentclass[twocolumn,prx,aps,longbibliography]{revtex4-1} 

\usepackage{color} 
\usepackage{amsmath}
\usepackage{amssymb} 
\usepackage{multirow}  
\usepackage{graphicx} 
\usepackage{soul}
\usepackage{url}
\usepackage{textcase}
\usepackage{bm}
\usepackage{epstopdf}
\usepackage{booktabs}
\usepackage{multirow}
\usepackage{makecell}
\usepackage[frenchb]{babel}
\usepackage[T1]{fontenc}
\usepackage{enumerate}

\newlength{\figurewidth} 
\ifdim\columnwidth<20cm 
\else 
\fi 
\definecolor{burntorange}{RGB}{191,87,0}
\definecolor{maskgray}{RGB}{200,200,200}

\newcommand{\rini}{r_\textrm{init}}
\newcommand{\rmax}{r_\textrm{max}}
\newcommand{\tpre}{t_\textrm{pre}}
\newcommand{\tpos}{t_\textrm{post}}
\newcommand{\etc}{\textit{etc.}}

\newcommand{\etal}{\textit{et al.}}
\AtBeginDocument{}

\begin{document} 

\title{On-Chart Success Dynamics of Popular Songs}
\author{Seungkyu Shin}
\affiliation{Graduate School of Culture Technology and BK21 Plus Postgraduate Programme for Content Science, Korea Advanced Institute of Science \& Technology, Daejeon, Republic of Korea 34141}
\author{Juyong Park}
\affiliation{Graduate School of Culture Technology and BK21 Plus Postgraduate Programme for Content Science, Korea Advanced Institute of Science \& Technology, Daejeon, Republic of Korea 34141}

\begin{abstract}
In the modern era where highly-commodified cultural products compete heavily for mass consumption, finding the principles behind the complex process of how successful, ``hit'' products emerge remains a vital scientific goal that requires an interdisciplinary approach. Here we present a framework for tracing the cycle of prosperity-and-decline of a product to find insights into influential and potent factors that determine its success. As a rapid, high-throughput indicator of the preference of the public, popularity charts have emerged as a useful information source for finding the market performance patterns of products over time, which we call the on-chart life trajectories that show how the products enter the chart, fare inside it, and eventually exit from it. We propose quantitative parameters to characterise a life trajectory, and analyse a large-scale data set of nearly $7\,000$ songs from Gaon Chart, a major weekly Korean Pop (K-Pop) chart that cover a span of six years. We find that a significant role is played by non-musical extrinsic factors such as the established fan base of the artist and the might of production companies in the on-chart success of songs, strongly indicative of the commodified nature of modern cultural products. We also review a possible mathematical model of this phenomenon, and discuss several nontrivial yet intriguing trajectories that we call the ``Late Bloomers'' and the ``Re-entrants'' that appears to be strongly driven by serendipitous exposure on mass media and the changes of seasons.
\end{abstract} 

\maketitle 

\section{Introduction} 
Competition for survival and success is a crucial mechanism underlying the evolution of species or actors in a complex system, be it biological, social, or technological~\cite{darwin2009origin,drossel2001biological,stuart1993origins}. The increasing availability of large-scale data along with a remarkable progress in the theory and modeling of complex systems has led to the emergence of the ``science of success'' that aims to reveal common patterns in the success of people or products in such diverse subjects as viral spreading of content, performance of athletes in sports, popularity of emergent technologies, and impact of scientific works~\cite{cintia2015network,garber2004density,griwodz1997long,sinatra2016quantifying}.

The scientific study of success is also deeply related to the tradition of developing robust and effective ranking methods for identifying the most successful and superior actors in a competition system~\cite{balinski2010majority,park2005network,price1976general,shin2014ranking,williams2000simple}.  Rankings of products and commodities serve a useful purpose for customers looking to purchase, or firms planning to advertise and market products. Cultural products such as popular songs are no exception, especially in this day and age where digital communication technology has enabled massive and efficient dissemination and consumption. Popularity charts have accordingly become more instantaneously updated, emerging as an essential reference for customers trying to make decisions in the face of a flood of new content and information, therefore becoming a coveted platform that affords products more exposure and prolonged success~\cite{burke1996dynamics,dixon1982lp,hesbacher1975sound,strobl2000dynamics}. This brings up the hope that we may now be able to search for answers to many interesting questions into the underlying mechanisms of successful products, including ``What are the features of `hit' products that can be learnt from their chart dynamics?'',  ``What are the factors--intrinsic or extrinsic--behind the success of products?'', and so forth.

In this paper, as an attempt to answer these questions using high-quality contemporary data and scientific methodology, we study the chart dynamics of K-Pop (Korean pop) songs, i.e. how the songs fare on popularity charts and what factors are behind it. K-Pop, a relatively recent international cultural sensation from South Korea, is characterised by catchy tunes and a heavy use of audiovisual elements.  One of the  early pinnacles of its global success happened in July of 2012 with PSY's \textit{Gangnam Style} that reached number two on the U.S. Billboard Hot 100.  Its online success can be said to be even more phenomenal, scoring nearly three billion views on YouTube as of this writing (April 2017) to become the most-watched online video.  Another prominent characteristic of K-Pop is the prominence of ``idols'', young, highly-trained dancer-singers who are designed to be the hub around which an entire industry of production and merchandise/service providers are organised~\cite{howard2006korean}.  This type of intensive commercialisation places K-Pop at the forefront of the contemporary music industry where the intrinsic, musical properties of a song such as its melody or lyrics are relegated to being merely one factor that affects its success~\cite{Adorno1941,hennion1983production,lopes1992innovation,negus1995mystical}. Prompted by this sweeping trend, in this paper we study a prominent K-Pop chart data to characterise the chart dynamics of successful songs, and then determine the extent to which such extrinsic factors correlated with their successes.

\section{Popularity Charts and The Life Trajectory of a Product}
\subsection{Data, methodology, and life trajectory patterns}

\begin{figure} 
\includegraphics[width=70mm]{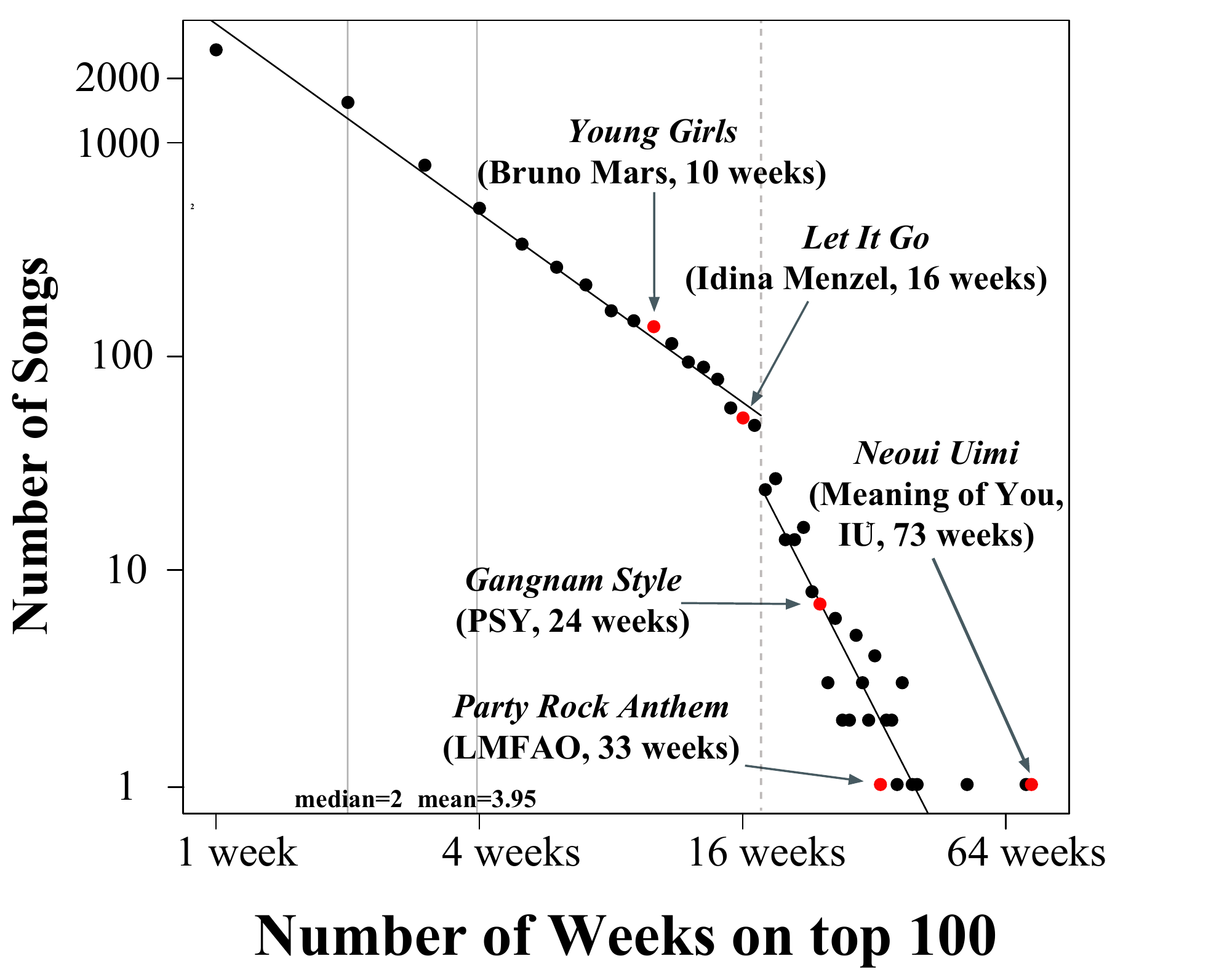} 
\caption{Histogram of the number of weeks spent by the songs on the Gaon Digital Chart that exhibits a highly skewed behavior. The mean is $3.95$ and the median is $2$ for the songs that spend at least one week on the chart. The longest is for IU's \textit{Neoui Uimi} (The Meaning of You) at $73$ weeks.} 
\label{figure01} 
\end{figure}

We analyze the data from Gaon Music Charts~\texttt{(www.gaonchart.co.kr)}, a collection of weekly music charts serviced by the association of Korea's music industry.  We focus on the Gaon Digital Chart, the signature one similar in spirit to the U.S. Billboard Hot 100.  It ranks the top-100 songs (both domestic K-Pop and foreign songs in a single chart) according to their digital sales figures including downloads, online streaming counts, and their offline album sales,~\etc~ (The exact weights given to each factor has not been made public, although the rankings themselves are open for anyone.) Our data covers all weeks between the second week of 2010 (the actual beginning of Gaon) and the 53rd week of 2015, for a total of 313 weeks.  During this period there have been in total $7\,560$ songs that appeared at least once on the chart.  The actual numbers of weeks spent on the chart, however, vary widely as shown in Fig.~\ref{figure01}: $36.4\%$  of the songs $(2\,750)$ appeared for one week only, whereas $9.1\%$ of the songs $(689)$ stayed on the chart more than ten weeks.  The longest life on the chart was enjoyed by IU's \textit{Neoui Uimi} (The Meaning of You) for a record of the $73$ weeks. The same artist's \textit{Joeun Nal} (Good Day) and PSY's \textit{Gangnam Style} topped the chart at number one for the most weeks $(5)$.

For most songs, the weekly rankings follow roughly a similar pattern (notable exceptions are discussed later): It appears in the chart, reaches its peak rank at some point, declines, and eventually leaves the chart.  Upon inspection of numerous curves made by the songs' rankings, we find it useful to characterise such \textbf{life trajectory} of a song via the following parameters:

\begin{center}
    \begin{tabular}{ | p{1cm} | p{5cm} |}
    \hline
    $\rini$	& Song's inaugural rank on the chart~\footnote{Since Gaon Digital Chart compiles weekly data starting on Sundays, the first visible rank of a songs can be significantly affected by at which point during the week it is released: A song released on a Monday, for instance, would have more time to accumulate stats (downloads and streaming counts) than another one released closer to a Sunday. To alleviate this problem, we actually use a song's second week rank as its inaugural rank $\rini$, reducing the number of songs analyzed to $4\,810$.}  \\ \hline  
    $\rmax$	& Peak rank  \\ \hline 
    $\tpre$ & Time taken to reach the peak since inaugural appearance on chart  \\ \hline 
    $\tpos$	& Time taken to exit from chart since peak rank\\ \hline 
    \end{tabular}
\end{center} 

\begin{figure} 
\includegraphics[width=80mm]{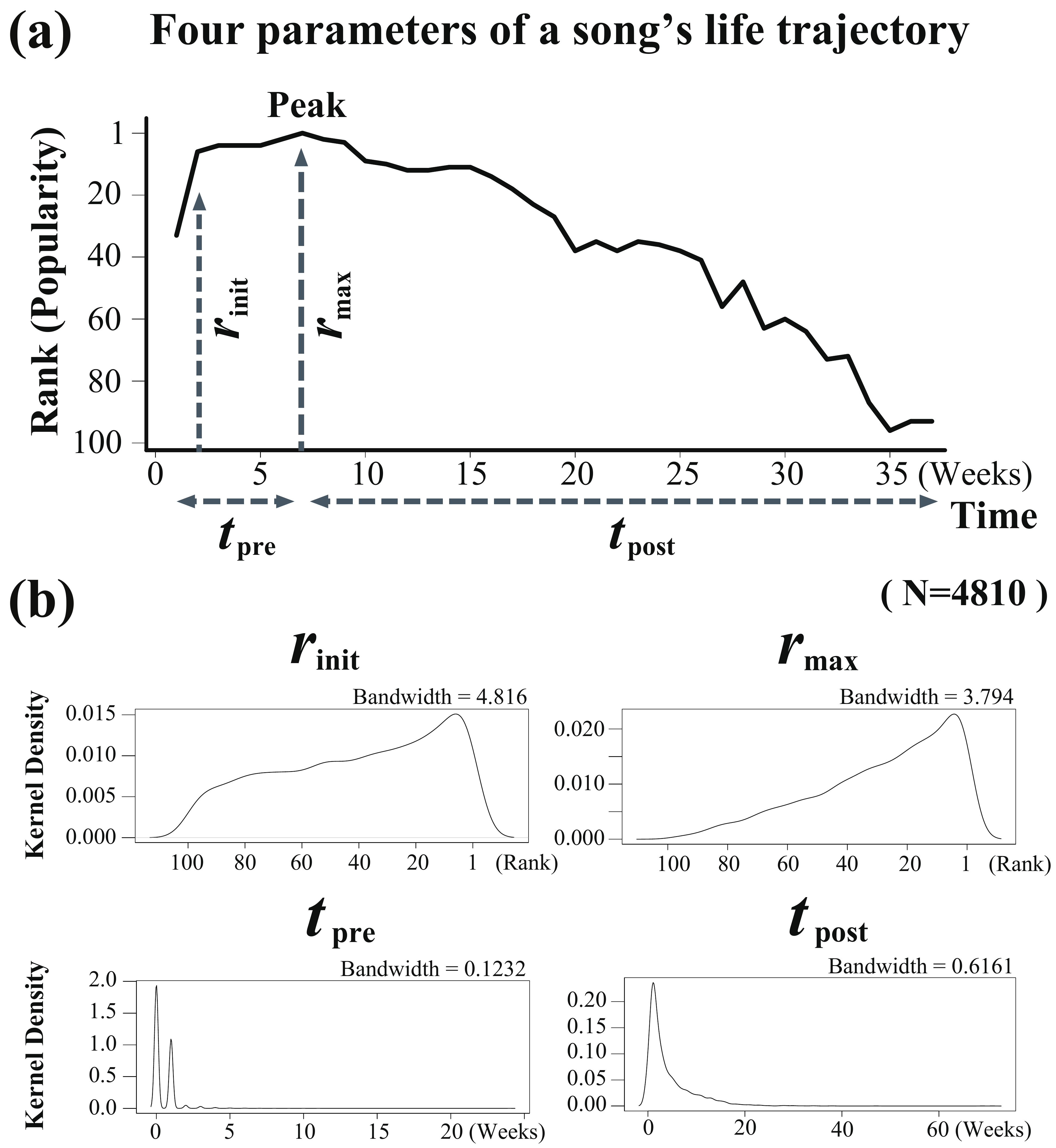} 
\caption{(a) Four parameters that describe a song's life trajectory on a chart: inaugural rank ($\rini$), peak rank ($\rmax$), time spent on the chart before peak rank ($\tpre$), and time spent on the chart after its peak ($\tpos$). (b) The distribution density of the four parameters estimated using Kernel Density Estimation (KDE) method. The $\rini$ and $\rmax$ are weighted towards high values, whereas $\tpre$ and $\tpos$ are weighted towards low values. Songs that appeared on the chart for two weeks or more ($4\,810$ songs in total, see definition of $\rini$ in Table.) were anlayzed to generate these figures.} 
\label{figure02} 
\end{figure}

An example is shown in Fig.~\ref{figure02}(a) for girlband EXID's \textit{Wiarae} (Up and Down). It first entered the chart at $\rini=7$, reaching its peak ($\rmax=1$)  $\tpre=6$ weeks later, then fell gradually (save for some brief rallies) for $\tpos=30$ weeks until it left the chart. In Fig.~\ref{figure02}(b) we show the probability density function of the four parameters using Kernel Density Estimation (KDE) method in the R statistical package~\cite{scott2015multivariate,sheather1991reliable}. $\rini$ and $\rmax$ are slightly skewed towards high values whereas $\tpre$ and $\tpos$ are more heavily skewed towards lower values, meaning that an overwhelming majority of songs enjoy only brief stints on the chart. We can also define the overall success of a song to be the area under its trajectory curve so that the longer-surviving, higher-ranking songs can naturally be called more successful.

\begin{figure} 
\includegraphics[width=60mm]{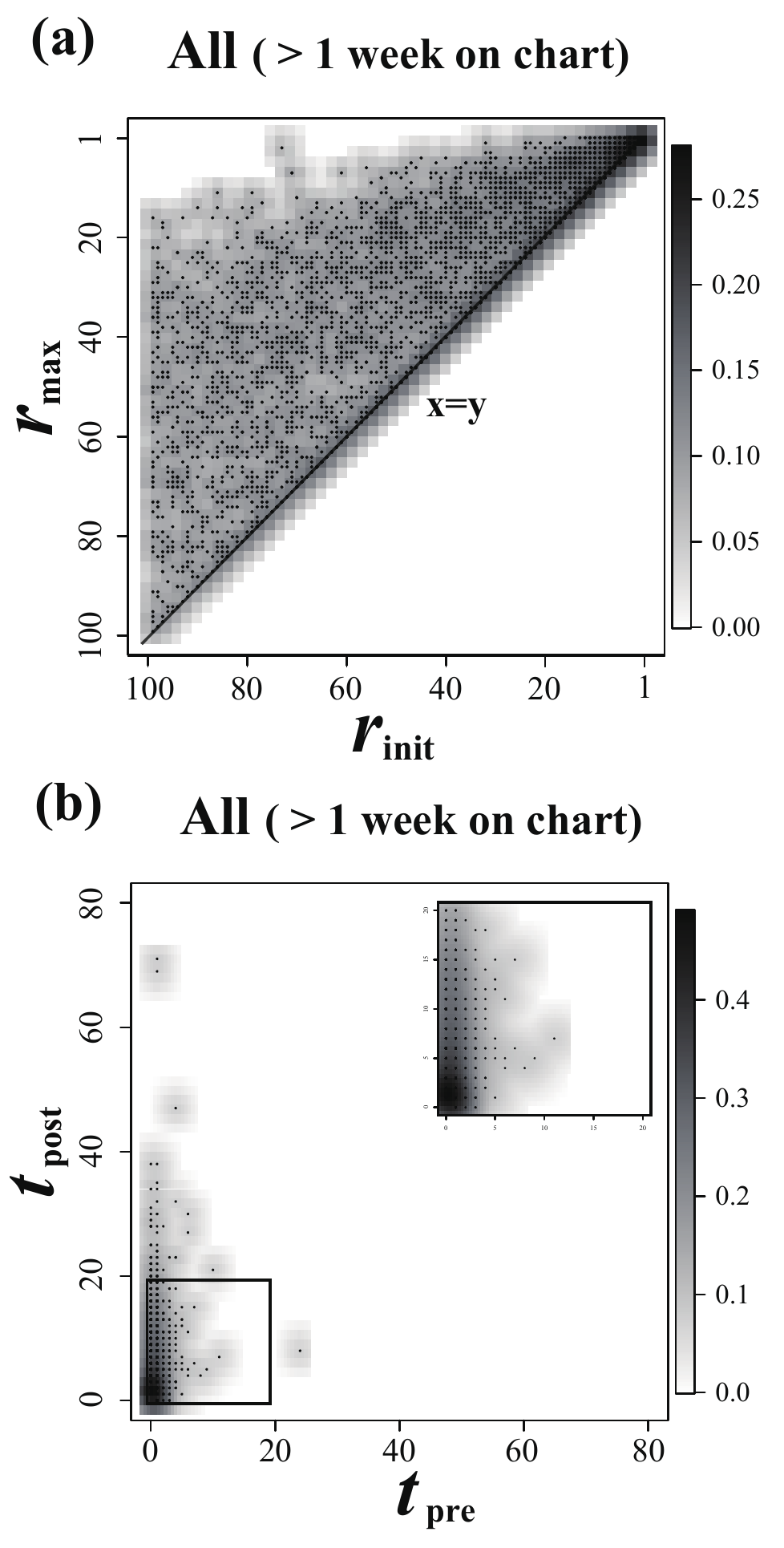} 
\caption{The life trajectory parameters of the songs on the chart. (a) The inaugural rank is plotted along the $x$-axis, and the peak rank is plotted along the $y$-axis. (b) The amount of time taken to reach the peak is plotted along the $x$-axis, and the amount of time taken to first go off the chart after reaching the peak is plotted along the $y$-axis.} 
\label{figure03} 
\end{figure}

\begin{figure*} 
\includegraphics[width=150mm]{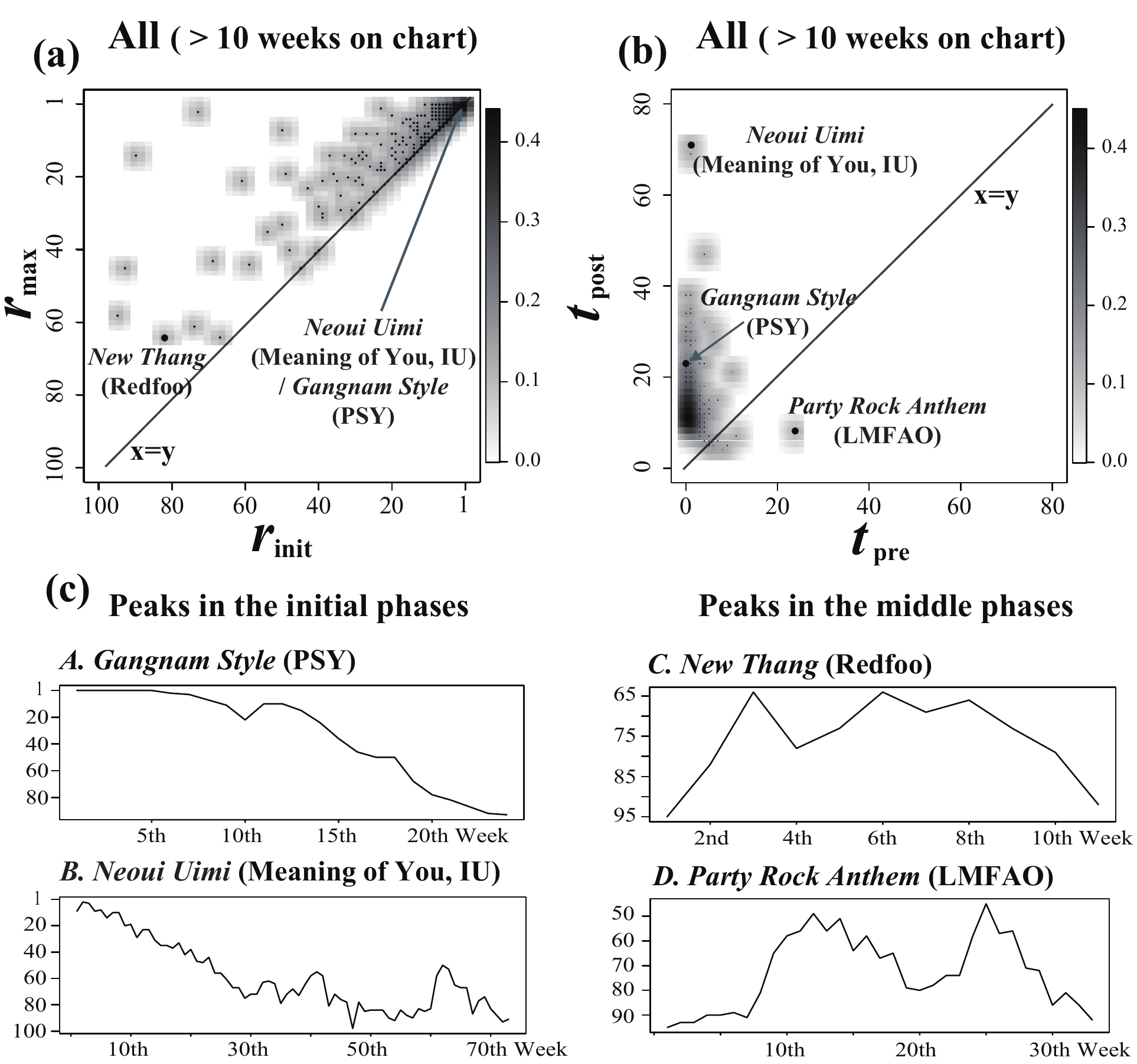} 
\caption{The life trajectory parameters of $689$ successful songs (defined as longer than ten weeks on the chart). (a) The bulk of the songs are located on the top right, meaning that a majority of songs that appear on the chart are ranked high in the early phases. (b) For most songs $\tpre$ ranges up to ten weeks while $\tpos$ is much more wide-ranging with a maximum of $71$ weeks.} 
\label{figure04} 
\end{figure*}

The relationships between the parameters reveal more interesting patterns about the trajectories of the songs,  presented in Fig.~\ref{figure03}~and~\ref{figure04}. In Fig.~\ref{figure03}(a), the inaugural ranks $\{\rini\}$ are plotted along the $x$-axis, and the maximum ranks $\{\rmax\}$ are plotted along the $y$-axis.  The data points appear to be roughly evenly spread out here with the exception of a high density of points along the diagonal  $\rini=\rmax$, meaning that for many songs the inaugural rank was indeed the peak. In Fig.~\ref{figure03}(b), the time until the peak $\{\tpre\}$ are plotted along the $x$-axis, and the time until exiting the chart since the peak $\{\tpos\}$ are plotted along the $y$-axis.  Since most songs stay on the chart for only a short period of time as shown in Fig.~\ref{figure01}, they are largely concentrated in the bottom left area.  But overall $\tpos>\tpre$, meaning that for most songs the decay is longer than ascension.

\begin{table*}[t]
\caption{Average trajectory parameters, and results of Kruskal test for 4810 songs (>1 week on chart).}
\label{table2}
\begin{center}
\begin{tabular}{c|c|c|ccccc|c}
    \toprule[1.2pt]
    \textbf{Variable} & \textbf{Subgroup} & \textbf{N} & \textbf{$\rini$} & \textbf{$\rmax$} & \textbf{$\tpre$} & \textbf{$\tpos$} & \textbf{Success} & \textbf{$p$-value}  \\
    \midrule[1.2pt]
    \multirow{3}{*}{Gender} & Male\textsuperscript{a}  & 2738  & 58.09  & 71.42  & 0.43  & 4.02  & 288.54 & \multirowcell{3}{All < 0.01** \\ but $\rmax$ : 0.31 }  \\
    & Female\textsuperscript{b} & 1684  & 60.74  & 72.41  & 0.48  & 4.38  & 320.90   \\
    & Mixed\textsuperscript{b}  & 388  & 61.52  & 72.60  & 0.51  & 4.55  & 327.91   \\
    \midrule
    \multirow{3}{*}{Type} & Solo\textsuperscript{a} & 2509  & 57.13  & 70.17  & 0.43  & 3.78  & 273.08 & \multirowcell{3}{All : 0.000*** \\ but $\tpre$ : 0.0025**}   \\
    & Group\textsuperscript{b} & 1897  & 61.70  & 74.12  & 0.46  & 4.69  & 339.11   \\
    & Collabo\textsuperscript{b}  & 404  & 61.49  & 71.76  & 0.55  & 4.38  & 319.80   \\
    \midrule
    \multirow{2}{*}{Nationality} & Domestic & 4654  & 60.05  & 72.70  & 0.43  & 4.19  & 305.79 & 
\multirowcell{2}{All : 0.000*** \\ but $\tpos$ : 0.12 }   \\
    & Foreign  & 156  & 36.88  & 46.74  & 0.96  & 4.03  & 221.19   \\
    \midrule
    \multirow{5}{*}{Genre} & Ballad\textsuperscript{a} & 1637 & 58.51 & 72.08 & 0.38 & 3.74 & 274.76 & \multirow{5}{*}{All : 0.000***}  \\
    & R\&B\textsuperscript{a} & 325 & 56.14 & 70.45 & 0.38 & 3.92 & 277.36    \\
    & Rock\textsuperscript{a}  & 311  & 55.05 & 72.43 & 0.30 & 3.91 & 276.42   \\
    & Dance\textsuperscript{b} & 1039  & 65.39 & 76.62 & 0.47 & 5.19 & 380.97   \\
    & Hiphop\textsuperscript{b}  & 653  & 61.05 & 70.95 & 0.48 & 4.31 & 308.15  \\
    \bottomrule
\end{tabular}
\end{center}
\end{table*}

In order to understand how these parameters are related to the extrinsic properties of the songs, we utilize additional metadata of the artists and the genres. The artist metadata consists of three properties: Gender (Male, Female, or Mixed), Type (Solo, Group, or Collaboration), and Nationality (Domestic or Foreign). The genre metadata consists of five major classes--Ballad, Dance, Rap and Hiphop, R\&B and Soul, and Rock--that each accounts for at least five percent of the songs in the entire data set. We use the analysis of variance (ANOVA) method to determine whether there exist statistically significant differences between groups of songs with those extrinsic properties. We compare the variance within groups and the variance between groups, then if the latter is significantly larger than the former, ANOVA lets us conclude that does exist a difference between the groups~\cite{hogg1995introduction}. The original ANOVA requires the population to be close to a normal distribution and population variances to be equal. For our kind of data that do not satisfy those conditions we use the Kruskal-Wallis test, non-parametric version of ANOVA~\cite{siegal1956nonparametric}. If the $p$-value from the test is less than or equal to the significance level ($0.05$), we reject the null hypothesis and conclude that not all populations are equal. In Table~\ref{table2}, we specify the average trajectory parameters of all groups based on the metadata and show the $p$-values from the Kruskal-Wallis test. The $p$-values shown in Table~\ref{table2} tell us that in most cases there exist statistically significant differences ($p<0.05$) depending on the properties of the songs. It is noteworthy that songs exhibit differences based on their extrinsic properties.  However, the significances are generally weaker for the Gender classes, and particularly $\rmax$ in regards to Gender ($p=0.31$) and $\tpos$ in regards to Nationality ($p=0.12$) show insignificant differences. This means that the artist's Gender is not a strong factor in a song's peak rank, and domestic (Korean) and foreign songs do not differ in the time they take to exit the chart.

To further characterize the differences between groups using ANOVA, we conducted follow-up tests to see whether certain  extrinsic properties led to different chart dynamics.  We see that ``Male'' and ``Solo'' artists show significantly lower values on all parameters in Table~\ref{table2}. Therefore their songs enter the chart lower, reach lower peak ranks, and exit the chart quickly. Amongst different genres, Ballad, R\&B and soul, and Rock do not show significant differences, while Dance and Hiphop genres are similar. Ballads are less successful than Dance and Hiphop music, reflecting the current state of the dance-heavy K-Pop.

When we study the more successful, ``hit'' songs, distinct patterns emerge. In Fig.~\ref{figure04}(a)~and~(b) we show analogous plots for the $689$ songs that stayed on the chart for longer than ten weeks, accounting for $9.1\%$ of the songs in the full data set. In Fig.~\ref{figure04}(a) we see that now the songs occupy the top right side, meaning that successful songs are likely to be already ranked high in its early phase, implying that a song's potential for on-chart success is realised at the very beginning, and appearing on the chart does not necessarily translate to opportunity for attaining a higher rank.  The message is similar in Fig.~\ref{figure04}(b) where $\tpre$ ($x$-axis) ranges merely between one and ten weeks (with the exception of \textit{Rock Party Anthem} by LMFAO which is not a K-Pop but a foreign song, which we will discuss later in more detail) while $\tpos$ on the $y$-axis assume much wider values with the maximum of 71 weeks. Consider PSY's \textit{Gangnam Style} and IU's \textit{Neoui Uimi} in Fig.~\ref{figure04}(c)~A~and~B that do show typical behaviors of K-Pop: Both take no to a very short time on the chart to reach their peaks.

\begin{table*}[t]
\caption{Average trajectory parameters, and results of Kruskal test for 689 songs (>10 weeks on chart).}
\label{table3}
\begin{center}
\begin{tabular}{c|c|c|ccccc|c}
    \toprule[1.2pt]
    \textbf{Variable} & \textbf{Subgroup} & \textbf{N} & \textbf{$\rini$} & \textbf{$\rmax$} & \textbf{$\tpre$} & \textbf{$\tpos$} & \textbf{Success} & \textbf{$p$-value}  \\
    \midrule[1.2pt]
    \multirow{3}{*}{Gender} & Male  & 361  & 91.63  & 94.43  & 0.88  & 13.93  & 934.56 & \multirowcell{3}{All > 0.05 \\ but $\rmax$ : 0.045* }  \\
    & Female & 259  & 93.32  & 95.35  & 0.88  & 13.41  & 953.95   \\
    & Mixed  & 69  & 93.48  & 96.57  & 0.86  & 14.41  & 972.33   \\
    \midrule
    \multirow{3}{*}{Type} & Solo & 274  & 91.72  & 94.38  & 0.85  & 13.94  & 940.89 & \multirow{3}{*}{All > 0.05}   \\
    & Group & 346  & 92.95  & 95.44  & 0.86  & 13.50  & 940.58   \\
    & Collabo  & 69  & 92.86  & 95.19  & 1.09  & 14.58  & 989.81   \\
    \midrule
    \multirow{2}{*}{Nationality} & Domestic & 669  & 93.34  & 95.64  & 0.80  & 13.74  & 947.25 & 
\multirowcell{2}{$\rini$, $\rmax$, $\tpre$ : 0.000*** \\ $\tpos$, Success > 0.05 }   \\
    & Foreign  & 20  & 62.65  & 73.45  & 3.60  & 15.45  & 891.60   \\
    \midrule
    \multirow{5}{*}{Genre} & Ballad & 182 & 94.51 & 96.13 & 0.72 & 14.02 & 960.70 & \multirowcell{5}{All > 0.05 }  \\
    & R\&B & 40 & 94.45 & 95.975 & 0.80 & 15.53 & 1070.18    \\
    & Rock  & 38  & 91.63 & 95.76 & 0.63 & 13.63 & 903.95   \\
    & Dance & 217  & 93.77 & 95.75 & 0.76 & 13.30 & 947.16   \\
    & Hiphop  & 106  & 93.85 & 95.71 & 0.71 & 14.27 & 953.06  \\
    \bottomrule
\end{tabular}
\end{center}
\end{table*}

We also find that foreign songs tend to show a noticeably different general behavior from that of K-Pop. Take \textit{New Thang} by Redfoo and \textit{Party Rock Anthem} by LMFAO, for instance, that are the outliers in Fig.~\ref{figure04}(a)~and~(b). \textit{New Thang} entered the chart at $\rini=83$, rising in the chart quite slowly (see Fig.~\ref{figure04}(c)~C), peaking at $\rmax=65$ and staying on the chart for a total of 11 weeks.  \textit{Party Rock Anthem}, on the other hand, took $\tpre=24$ weeks to peak, but exited the chart in $\tpos=8$ weeks.  In fact, the other outliers in the plots of Fig.~\ref{figure04}(a)~and~(b) tend to be foreign songs as well, taking a longer time to reach their peaks than K-Pop songs: Six out of eight songs below diagonal $x+y=100$ in Fig.~\ref{figure04}(a) are foreign songs, and below $y=x$ in Fig.~\ref{figure04}(b) (meaning  $\tpre>\tpos$), four out of the seven songs were foreign although only $5.41\%$ $(260)$ of the songs analyzed ($4\,810$) are foreign.

The Kruskal-Wallis test also reveals significant differences between domestic and foreign songs with the group of $689$ songs that stayed on the chart for more than ten weeks. The average trajectory parameters of these groups and $p$-values from the Kruskal-Wallis tests are given in Table~\ref{table3}.  Here, unlike the entire group of songs, artist features and genres do not show significant differences ($p>0.05$). However, domestic and foreign artist groups exhibit significant differences regarding $\rini$, $\rmax$, and $\tpre$ ($p<0.0001$).  Although the overall success are not significantly different, foreign songs have lower $\rini$ and $\rmax$, and take longer times to reach their peaks. These observations regarding hit songs and their artist properties imply that while non-musical extrinsic properties of popular songs do in general matter for songs' on-chart behaviors, their effects become weaker when successful songs are considered and the intrinsic, musical qualities of the songs become more important.

\begin{figure*} 
\includegraphics[width=150mm]{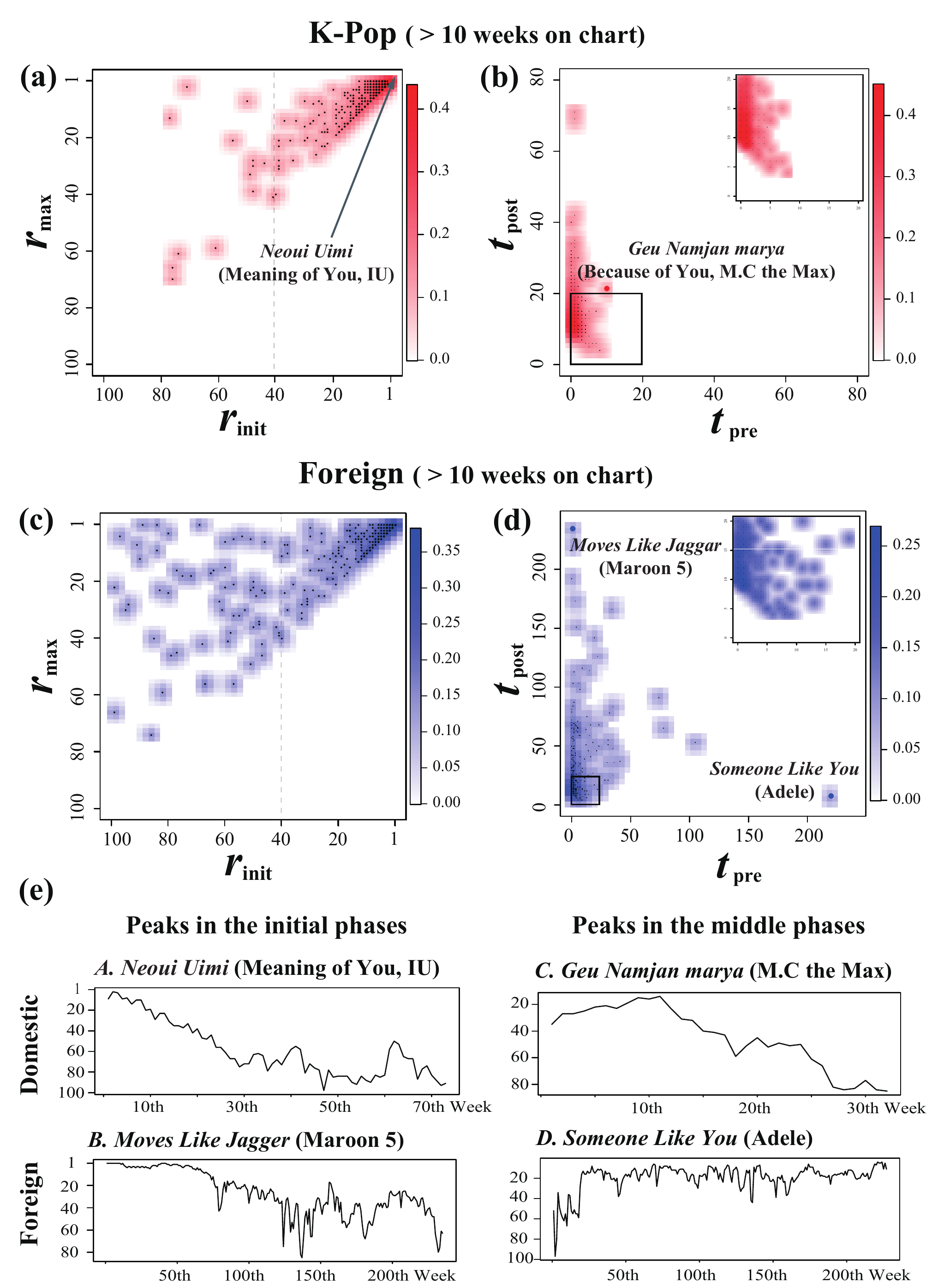} 
\caption{The analoguous scatter plots for K-Pop (domestic) songs (red) and foreign songs (blue). Foreign songs are comparatively more spread out, probably due to less influence of extrinsic factors.} 
\label{figure05} 
\end{figure*}

\subsection{Chart dynamics of K-Pop vs foreign songs} 
We now take a deeper look at the differences between K-Pop and foreign songs. To do so we analyze two separate subcharts of Gaon, primarily due to the small sample size of foreign songs in the original data: Of the $N=7\,560$ songs that ever appeared in the main Gaon digital chart, only three percent $(260)$ are foreign.  The subcharts we use here are the digital top-100 charts exclusively for K-Pop and foreign songs.  This yields a much larger set of foreign songs for us to analyze: $7\,602$ for K-Pop (there is not much difference, since the main chart was dominated by domestic songs to begin with), and $3\,855$ for foreign songs. Again focusing on those that survived for more than ten weeks on each chart ($702$ K-Pop and $327$ foreign songs), we find that the total life on the chart already show significant differences: K-Pop songs survive on average $15.7\pm0.4$ weeks with maximum 73 weeks by IU's \textit{Neoui Uimi}, while foreign songs survive $37.3\pm1.0$ with a maximum of 235 weeks by Maroon 5's \textit{Moves Like Jagger} (Fig.~\ref{figure05}(e)~B).

Their life trajectories are plotted separately in Figs.~\ref{figure05}(a)~to~(d). They again confirm that K-Pop songs reach their peaks early, with only $2.14\%$ entering the chart under the 40th place, and the longest $\tpre$ is equal to ten, for \textit{Geu namjan marya} (Because of You) by M.C the Max. Foreign songs on the other hand, while fewer than half the number of K-Pop in total, show a wider distribution of the parameters in Fig.~\ref{figure04}(c)~and~(d):
 $18.7\%$ entered the chart ranked 40th or lower, and $46$ songs took more than ten weeks to reach their peaks, with maximum $\tpre=220$ by Adele's \textit{Someone Like You} that entered the chart at $\rini=98$.

The question is, then, which factors are behind such visible differences between K-Pop and foreign songs. One could argue that it could be the music itself, if there really are intrinsic differences between K-Pop and foreign pop. Even if such differences from a musical standpoint did exist at all, however, they would be quite subtle, given that nearly all modern popular genres--rock and roll, ballads, synth pop, electronica,~\etc--in foreign pop are also well represented in K-Pop. We therefore find it unconvincing that the musical differences would be the sole determinant of the differences in the chart dynamics of songs.  If not the intrinsic properties, then some external factors may be in play here. While there could be many, here we study two that are widely believed to be defining characteristics of contemporary K-Pop, the artists and the machinery of production companies.

\section{Artists, Production Companies of K-Pop} 
The observed fact that most songs reach their peaks in the very early phases of their lives does seem unnatural and counterintuitive--this means that the songs have already realised potential for success during the one short week before entering the chart or, even more bizzare, before its release. We believe that the clues to the origin of such behavior comes from a dominant recent trend in how new K-Pop songs are produced and marketed: Production companies are increasingly focusing on pre-release marketing to rally the fandom that prop up online streaming counts and downloads in the very early stages of the song's release~\cite{howard2006korean}. This happens in tandem with the actions of the fan base of the artist who are not passive consumers in music industry but active disseminators of the songs to the general public outside the fan base. This means that production companies in K-Pop--influential entities that recruit, train, finance, manage, and market artist--and huge fan bases that reflect the artists' prior success and reputation may hold significant sway over the chart success of new songs~\cite{peterson1971entrepreneurship,peterson1975cycles}, possibly even more than their qualities may. Artists and producers of foreign pop songs, on the other hand, generally do not maintain such a strong presence in Korea which explains their more natural dynamics on the charts (Fig.~\ref{figure04}(e)~and~(f)).   

We now investigate the influence of artists and production companies on a song's chart dynamics. We start by defining a quantity that represents the influence of an artist and a production company. Possible candidates could include the size of the fan base, the company's sales volume, stock price, and number of affiliated artists in companies. But here we focus on chart performance as the measure of an artist and a company's influence.  In other words, we posit that a history of chart success is an indicator of the entity's influence and future success. This is inspired by the so-called ``Matthew effect'', the rich-get-richer phenomena often observed in social systems~\cite{merton1968matthew}. We define the \textbf{Chart Success Index} $S_i$ of an artist and a company $i$ as the sum total of the weekly ranks of all songs produced by the subject in the history of Gaon Digital Chart, i.e.
\begin{equation} 
	S_i\equiv\sum_{\substack{
					w\in\textrm{All Weeks}\\
					s\in\Omega^w_i}}
					\biggl(101-r^w_s\biggr),
\end{equation}
where $\Omega^w_i=\{s\}$ refers to all songs from artist or company $i$ ranked on the chart on week $w$, and $r^w_s$ refers to the weekly rank of the song.  

\begin{figure*} 
\includegraphics[width=140mm]{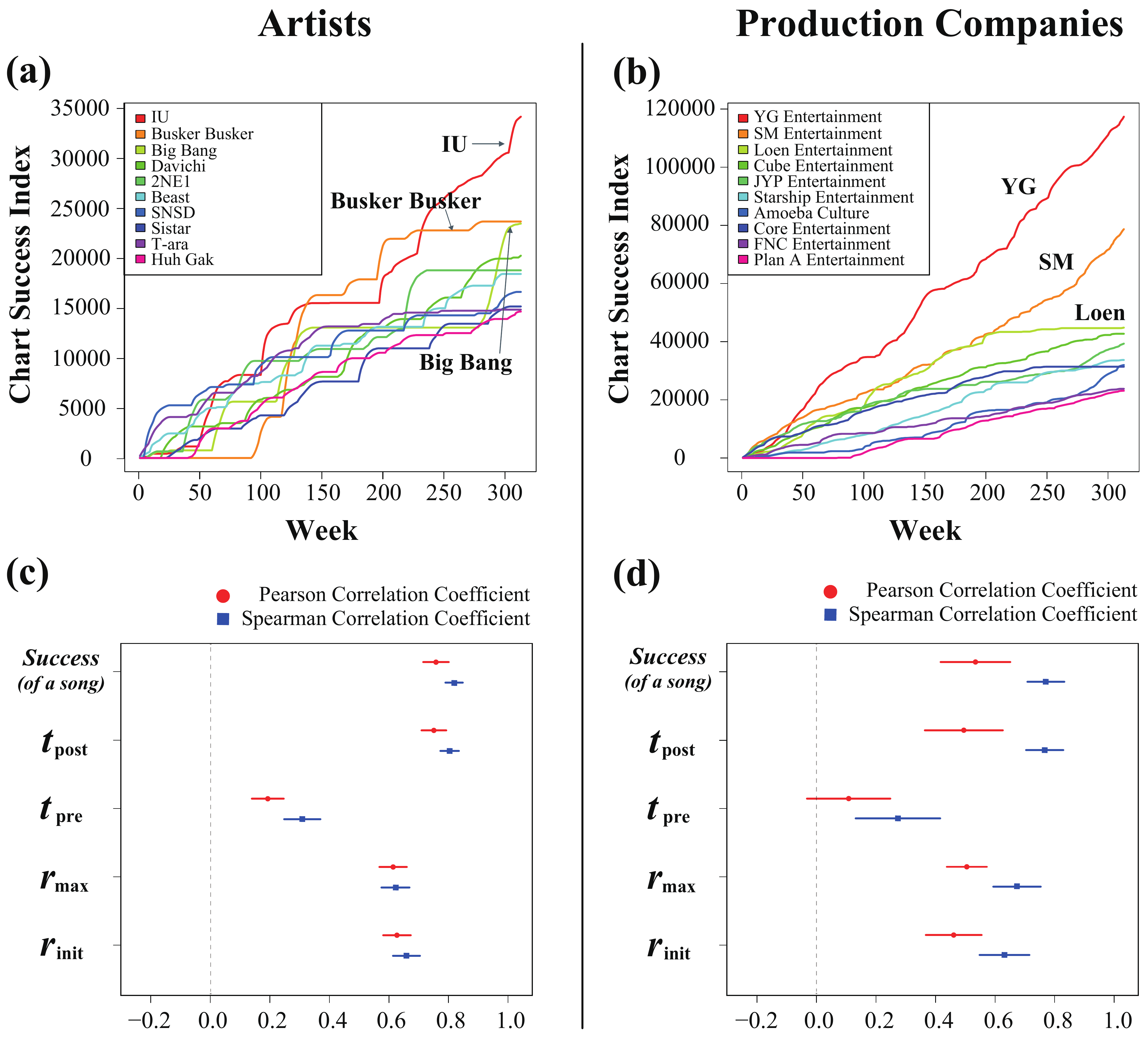} 
\caption{(a)--(b) The growth of chart success index $S$ of top 10 artists and production companies. (c)--(d) Correlation between the chart success index and average life trajectory parameters of song released by the artists and the production companies. The errors are estimated using the jackknife method~\cite{newman1999monte}. All parameters except $\tpre$ are positively correlated.} 
\label{figure06} 
\end{figure*} 

First, Fig.~\ref{figure06}(a) shows the $S$ values of the ten most successful artists by the last week in our data set.  Most artists display a pattern of alternating rises and plateaus in the growth of $S$, the plateaus meaning the time between successive songs. While it is the band that generally occupy the top positions, IU is a notable exception.  Second, Fig.~\ref{figure06}(b) shows the $S$ values of the top ten production companies. Most companies display linear growth patterns unlike their artist, with YG, SM, and Loen Entertainment corporations being the major three as of the last week in our data. To view the relationship between an entity's influence and the life trajectory of the songs they release, we compute the correlation  between final value of $S$ and the four curve parameters plus overall success of each song (the area under its trajectory curve). The errors are estimated using the jackknife method~\cite{efron1979computers,newman1999monte}. Here we note that we consider the life trajectories of songs from \emph{debuting} artists, since unknowns in the market their production companies' abilities and clout would be the only major external, non-musical factor behind their chart performance. Of the $934$ production companies that produced at least one new artist on the chart, $514$ produced only one. To eliminate similar errors originating from such small players, we consider $193$ artists (out of $1\,680$) that put more than ten songs, and the $55$ companies that put nine or more debuting artists on the chart. The results are shown in Fig.~\ref{figure06}(c)~and~(d).  

All parameters but the pre-peak time ($\tpre$) show strong positive correlations with $S$. This means that being a renowned artist or having powerful production company behind one's back is highly correlated with general chart success (i.e. long duration on chart and high peak rank), but regardless of the prior reputation or company's influence the artist hits their peak early.  To a skeptical eye, this phenomenon may be a reason for doubting the role of the ``quality'' of a K-Pop song in its chart performance; since early peaking (short $\tpre$) is universal while $\rmax$ is positively correlated with $S$, $\rmax$ could be due simply to the marketing and the established fan base of a company--because it suggests that a K-Pop song's success was somewhat determined even before its release, before it had a chance to be introduced to the larger consumer base--and $\tpos$ would also be a straightforward consequence of that (the higher the rank, the more time it takes to leave the chart). It should be noted, however, that this is not a complete picture, and there is still evidence that quality and musical properties do matter in a song's chart success.  Ironically, one could make the point from the dynamics of foreign songs in the Korean market that lack such a level of support and marketing from production companies: Foreign songs are slow to gain popularity, but they tend to have more staying power.  It may be an unsatisfactory state of affairs for the business that the effect of marketing may eclipse the importance of musical quality for K-Pop, and it remains to be seen how much role quality plays in the long run~\cite{henard2014all,mauch2015evolution,ni2011hit}. 

\section{Endogenous versus Exogenous Peaks}

Our analyses on the the impact of extrinsic properties separate from the intrinsic ones of popular songs are in line with earlier studies on the two types of dynamical features in complex networks: exogeneity and endogeneity~\cite{kubo1957statistical,albeverio2006extreme,holyst2000phase} that are believed to cause critical peaks. Exogenous peaks occur in response to external kicks, while endogenous peaks are spontaneous internal fluctuation formalized by the theory of self-organized criticality. It is therefore well understood that it is important to distinguish between exogeneity and endogeneity to measure their respective effects to ultimately understand the fundamental dynamics underneath a system. In particular, Sornette~\etal~\cite{sornette2004endogenous} introduced a model of epidemic propagation which specifies that exogenous peaks occur abruptly and are followed by a power-law relaxation while endogenous peaks are characterized by approximately symmetrical power law growth and relaxation. According to the model, the endogenous peaks ascend and descend slowly proportional to $1/|t-t_c|^{1-2\theta}$ where $t=0$ corresponds to the peak maximum and $t_c$ is an additional positive parameter. On the other hand, the relaxation of exogenous peaks behave like $1/(t-t_c)^{1-\theta}$ which decays much faster than endogenous peaks. Sornette also found an average value of the exponent $\theta\sim0.3 \pm 0.1$.  However, Lambiotte~\cite{lambiotte2006endo} observed that the relaxation process can be as well seen as an exponential one that saturates toward an asymptotic state when focusing on the short-time scale after a sales maximum ($\sim$ 1 month). He suggested a thermodynamic model to represent the decays with a relaxation coefficient $\lambda$ which is inversely proportional to the relaxation time, and found that the initial decay of exogenous and endogenous peaks occurs on different time ranges: $<\lambda>_{exo}\sim2<\lambda>_{endo}=0.14$.

\begin{figure} 
\includegraphics[width=70mm]{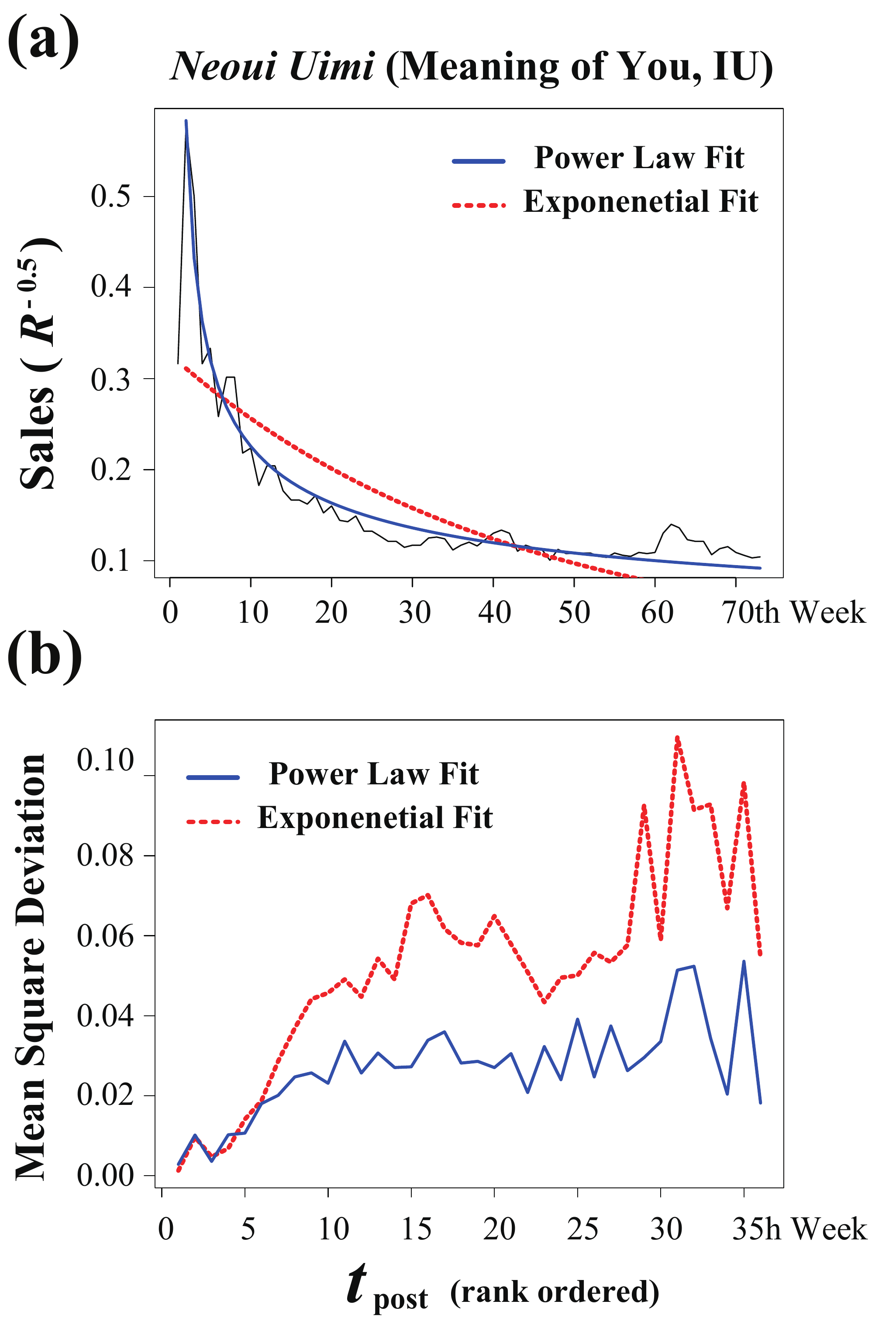} 
\caption{(a) Power-law (blue) and exponential (red) fits to IU's \textit{Neoui Uimi}, the longest-surviving song on the chart. The power law-fit appears better matched. (b) The change of root-mean-square deviation along the relaxation time, $\tpos$. At smaller $\tpos$, the exponential fit (red) shows similar or better results than the power-law fit (blue), but the power-law fit shows smaller residual errors at large $\tpos$.} 
\label{figure07} 
\end{figure} 
 
We apply these models to our data, limited to the $689$ songs that stayed on the chart more than ten weeks for sufficient data points. First, we transform a time series of the rank $R$ of a song into a time series of instantaneous sales flux through the relation $S=R^{-\gamma}$, $\gamma=0.5$ following Sornette~\etal's approach. We fit the relaxation of life trajectories with power-law and exponential models respectively, and measure the root mean square deviation in order to quantify the goodness of fit. Lower values of the deviation imply a better fit. Fig.~\ref{figure07}~(a) shows the power-law and exponential fits of the song \textit{Neoui Uimi}, the longest-ranked on the chart, and Fig.~\ref{figure07}~(b) shows the changes in the root-mean-square deviation along the relaxation time, $\tpos$. At smaller $\tpos$ ($<5$ weeks), the exponential fit (red) showed similar or better results than the power-law fit (blue), but the power law seems to have lower residual errors at large $\tpos$, implying that the relaxation process behave exponentially at the shorter time scale.

\begin{figure} 
\includegraphics[width=70mm]{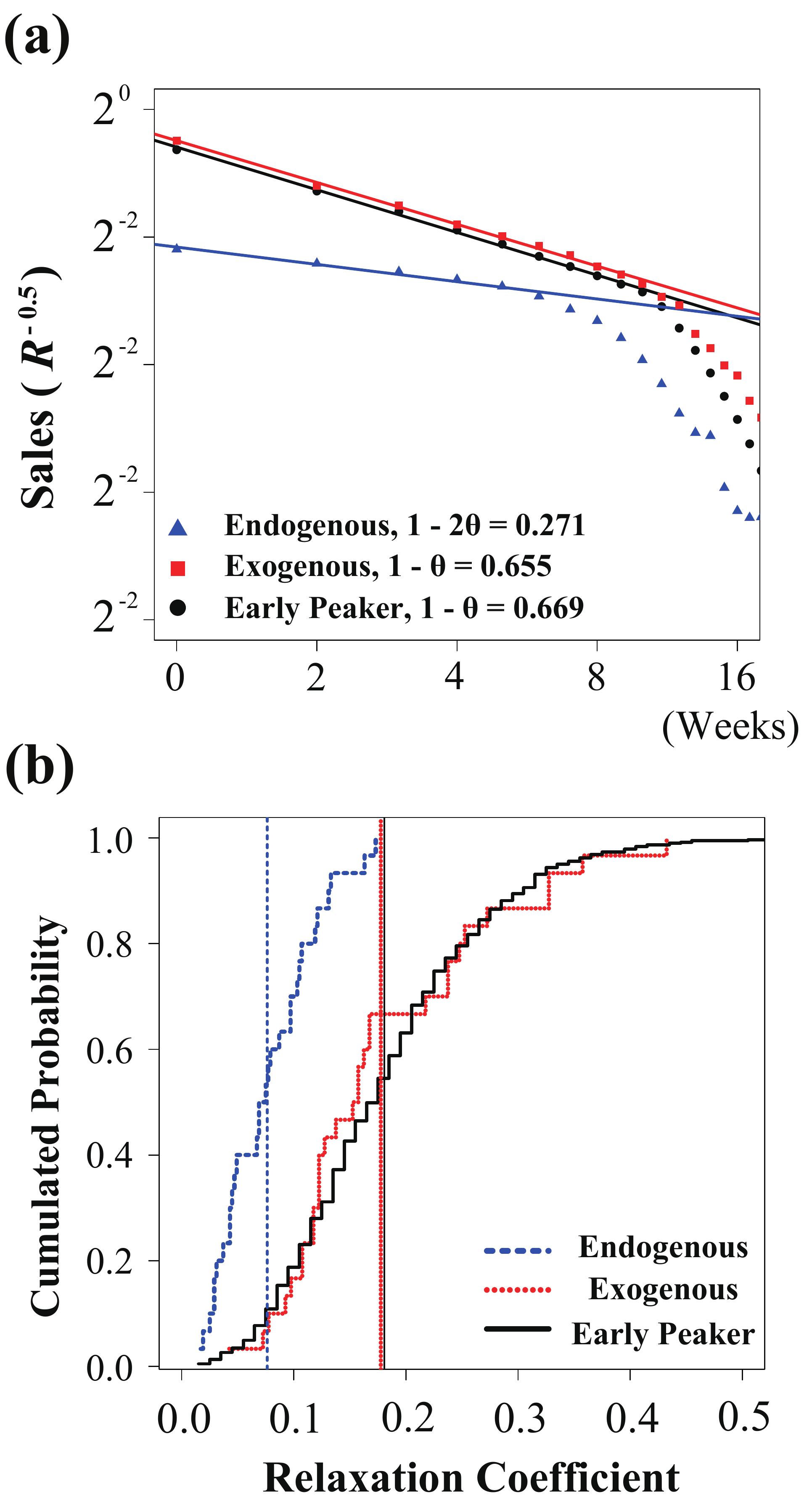} 
\caption{(a) Relaxation of sales after peaks obtained by averaging over songs in each class classified according to the precursory acceleration. (b) The cumulated histograms of relaxation coefficient $\lambda$ for three classes: endogenous (blue), exogenous (red), and early peaks (black). The vertical lines point to the average values $0.076$ (endogenous), $0.178$ (exogenous) and $0.181$ (early peaks). The exogenous and endogenous classes are markedly distinguishable, and early peak group shows a similar behavior to exogenous peaks.} 
\label{figure08} 
\end{figure} 
 
In addition, we extract endogenous and exogenous peaks inferred from their initial acceleration defined as $\frac{\rmax^{-0.5}-\rini^{-0.5}}{\tpre}$ here based on the assumption that exogenous peaks occur abruptly while endogenous peaks are formed relatively slowly. 30 high-ranked and low-ranked peaks are respectively classified as exogenous and endogenous. Meanwhile, as most songs ($607$ out of $689$ songs) reach their peaks immediately as they are released, the initial acceleration is not defined ($\rmax-\rini=0$, $\tpre=0$), so we classify them as a separate ``Early Peaker'' group to observe whether they show behaviors similar to those of endogenous or exogenous peaks. Fig.~\ref{figure08}~(a) shows relaxation of sales after peaks obtained by averaging over songs in each class, and Fig.~\ref{figure08}(b) present the cumulated probability of the relaxation coefficient $\lambda$ in the three classes. Both figures clearly indicate difference between endogenous and exogenous peaks being separated significantly. The power law exponents in Fig.~\ref{figure08}~(a) also approximately obey Sornette~\etal's prediction, $1-\theta$, and $1-2\theta$. In addition, the average relaxation coefficient of exogenous peaks ($0.178$) is over twice that of endogenous peaks ($0.076$), implying that the relaxation time is significantly shorter in exogenous shocks. Interestingly, the early peaker songs without any precursory pattern also show a similar behavior to exogenous peaks in both the power law exponent and relaxation coefficients, leading us to infer that their early peaks can indeed be attributed to external forces rather than spontaneous fluctuations.

\section{Discussions}  
In this paper we have studied the life trajectories of K-Pop and foreign songs on Korea's Gaon Charts.  We have shown that a large majority of life trajectories could be represented with four parameters of their peak rankings and duration on the charts. Many songs, especially K-Pop, were found to attain their peaks in the early phases, which indicated the influence of non-musical external factors such as the power of artists and production companies.  

We have demonstrated the possibility of utilizing popularity data to describe how the state of an industry affects the success of products, and we believe that the application is not necessarily limited to popular songs. We have explored only a small range of possibilities, and we hope to see our methods applied to more systems in a wider gamut of products and markets. 

\begin{figure} 
\includegraphics[width=70mm]{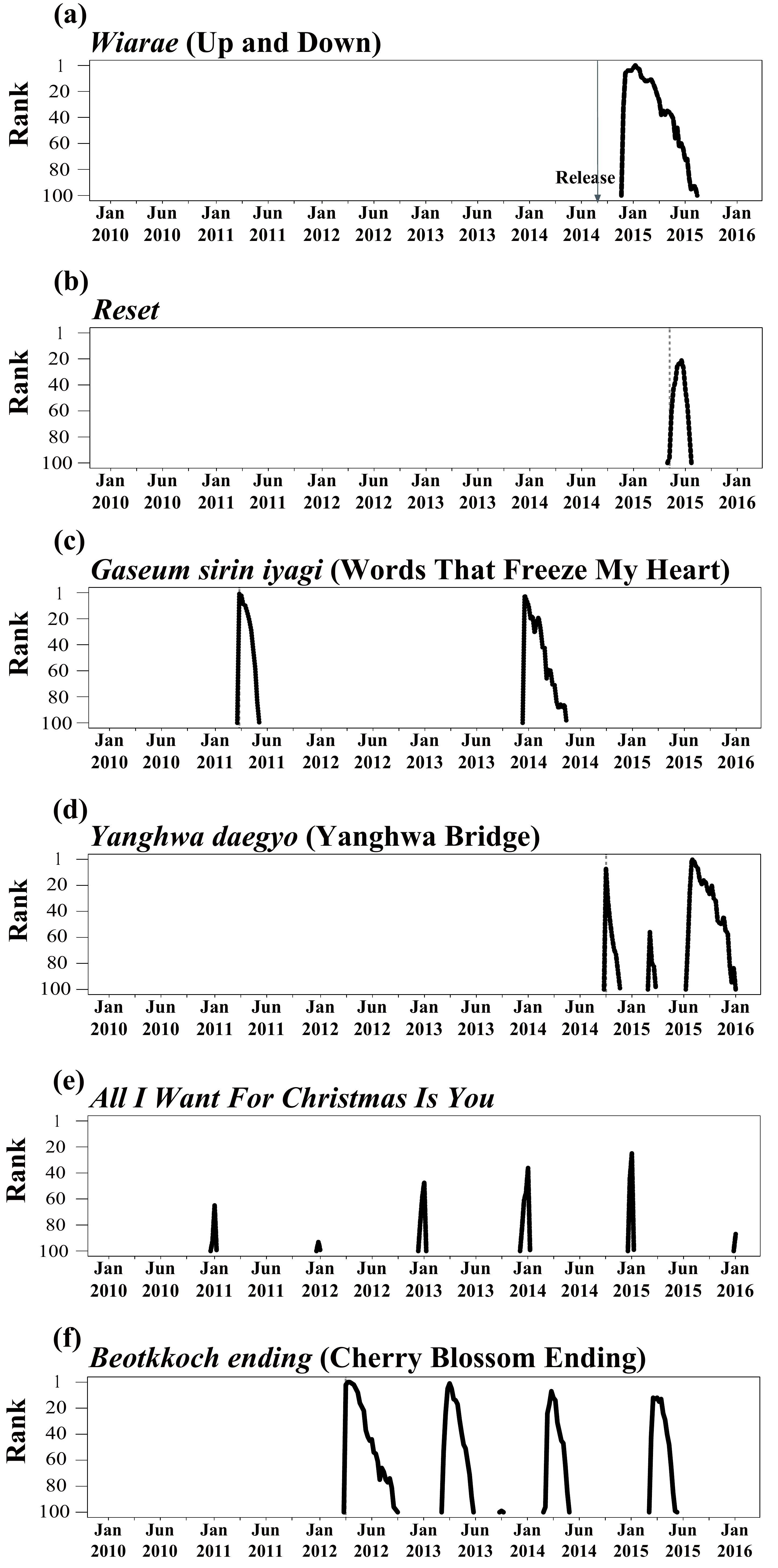} 
\caption{Two types of notable outlier types observed in chart dynamics. (a)--(b) The first type are ``Late Bloomers,'' referring to songs that climb up the chart or peak noticeably later than their market releases, often due to belated media exposure. (c)--(f) The second type are ``Re-entrants,'' referring to the songs that re-enter a chart after leaving it. Media exposure and seasonal changes are the main factors.} 
\label{figure10} 
\end{figure}  

We now conclude this paper by introducing two classes of songs that do not follow the general trends of Fig.~\ref{figure04} but could still be interesting. The first class is ``Late Bloomers'' that take an relatively long time since debut to climb up the chart. This would be a very rare feat indeed: Only $1.5\%$ of the songs (112 in total) manage to climb in rankings for three straight weeks.  A good example is \textit{Wi-arae} (Up and Down) by EXID: having failed to make it to Gaon upon release, it went viral on other social networking sites before taking the 1st place several weeks later, as shown in Fig.~\ref{figure10}(a). Another example is \textit{Reset} by Korean-American rapper Tiger JK.  The song was intended as a soundtrack to a TV drama whose popularity pushed up the song on to the charts, and its peak in the middle of its trajectory matches with the ratings peak for the TV show, see Fig.~\ref{figure10}(b). The second class of extraordinary patterns is ``Re-entrants'', referring to the songs that return to the chart after falling from it the first time (we consider a song's life trajectory on the chart has come to an end when it exits the chart and does not return within five weeks.). A very small percentage of the songs ($1.5\%$, or 116 songs) are such cases, and an even smaller number $(39)$ stay on the chart for longer than five weeks after re-entry. A detailed review of the forces behind such behavior tells us that there are broadly two kinds: The first is renewed media exposure and broadcasting. Particularly, getting featured on audition programs such as \emph{K-Pop Star} (Korean version of American Idol) or game shows centered on songs is a common way by which an old song experiences a surge of interest. \textit{Gaseum Sirin Iyagi} (Words That Freeze My Heart) by Wheesung (Fig.~\ref{figure10}(c)) and \textit{Yanghwa daegyo} (Yanghwa Bridge) by Zion.T in Fig.~\ref{figure10}(d) are famous examples helped by the mass media, sometime gaining even more popularity after re-entry than upon its release. The second is seasonal effect, which is very straightforward: Christmas carols such as \textit{All I Want For Christmas Is You} by Mariah Carey (Fig.~\ref{figure10}(e)) is a good example. The K-Pop song \textit{Beotkkoch Ending} (Cherry Blossoms Ending) by Busker Busker, popular upon its release, is an example that became a spring carol of sorts that re-enters the chart every spring in Korea, as seen in Fig.~\ref{figure10}(f). To conclude, there were evidence that, absent powerful production companies, quality and other factors such as media exposure could lead to chart success resulting in extraordinary life trajectories such as the late bloomers and the re-entrants.

\section*{Acknowledgments}
This work was supported by the BK21 Plus Postgraduate Organisation for Content Science,  IITP grant funded by the Korean government (MSIP-R0115-16-1006), and National Research Foundation of Korea (NRF-20100004910, NRF-2016S1A2A2911945, and NRF-2016S1A3A2925033).

\bibliography{bib_Kpop}  
 
\end{document}